\documentclass[11pt]{article}
\usepackage{graphicx}
\usepackage{dcolumn}
\usepackage{amssymb}
\usepackage{bm}
\usepackage{enumerate}

\def\bdoc{
\begin{document}} \def\edoc{\end{document}}

\newcounter{item}
\def\beq{\begin{equation}} 
\def\eeq{\end{equation}}
\def\bea{\begin{eqnarray}} 
\def\eea{\end{eqnarray}}
\def\ben{\begin{enumerate}}
\def\een{\end{enumerate}} 
\def\bitem{\begin{itemize}}
\def\eitem{\end{itemize}}
\def\la{\langle}
\def\ra{\rangle} 

\def\a{\alpha} 
\def\b{\beta} 
\def\g{\gamma} \def\G{\Gamma} 
\def\d{\delta} \def\D{\Delta} 
\def\e{\epsilon} \def\vare{\varepsilon} 
\def\k{\kappa}
\def\l{\lambda}\def\L{\Lambda}
\def\m{\mu}
\def\n{\nu}
\def\o{\omega}\def\O{\Omega} 
\def\p{\pi}\def\P{\Pi}
\def\r{\rho}
\def\s{\sigma} \def\S{\Sigma}
\def\t{\tau}
\def\th{\theta}\def\Th{\Theta}
\def\x{\xi}
\def\z{\zeta}
\def\bfs{\mbox{\boldmath$\sigma$\unboldmath}}
\def\bfp{\mbox{\boldmath$p$\unboldmath}}
\def\bfA{\mbox{\boldmath$A$\unboldmath}} 
\def\bfsp{\bfs\! \cdot\!\bfp} \def\half{\textstyle{\frac{1}{2}}}
\def\3halfs{\textstyle{\frac{3}{2}}} \def\em{\it}

\def\bfE{{\bf E}}
\def\bfB{{\bf B}}
\def\bfj{{\bf j}}
\def\bfS{{\bf S}}
\def\curlE{\bfn\times\bfE}
\def\curlB{\bfn\times\bfB}
\def\divE{\bfn\cdot\bfE}
\def\divB{\bfn\cdot\bfB}
\def\divj{\bfn\cdot\bfj}
\def\dt{\partial_{t}}
\def\dz{\partial_{z}}
\def\eminv{\frac{1}{\epsilon_0\mu_0}}
\def\bfn{{\bf\nabla}}

\def\e{\mbox{\boldmath$\epsilon$\unboldmath}}
\def\r{\mbox{\boldmath$\widehat{x}$\unboldmath}}
\def\bfs{\mbox{\boldmath$\sigma$\unboldmath}}
\def\bfn{\mbox{\boldmath$\nabla$\unboldmath}} 
\def\bfn{{\bf\nabla}} 
\def\bfk{{\bf k}} 
\def\bfr{{\bf r}} 
\def\vece{\vec{\epsilon}}
\def\bfs{\vec{\sigma}} 
\def\scri{${\cal I}^+$ }
\def\scriminusi{${\cal I}^-$ }
\bdoc

\begin{center}
 {\LARGE \bf Horizon Entropy}\\
\vskip 5mm
Ted Jacobson\\
{\it Department of Physics,
University of Maryland\\
College Park, Maryland 20742-4111}\\
\vskip .3 truecm
Renaud Parentani\\
{\it Laboratoire de Math\'ematiques et Physique Th\'eorique, UMR 6083\\
Universit\'e de Tours, 37200, France}\\
\end{center}

\abstract{
Although the laws of thermodynamics are 
well established 
for black hole horizons, much less has been said in the literature to 
support the extension of these laws 
to more general settings such as an asymptotic de Sitter horizon or
a Rindler horizon (the event horizon of an asymptotic 
uniformly accelerated observer).  
In the present paper we review the results that have been previously
established and argue that the laws of black hole thermodynamics, 
as well as their underlying statistical mechanical content,
extend quite generally to what we call here ``causal horizons". 
The root of this generalization is the 
local notion of horizon entropy density.
}

\section{\large INTRODUCTION}

Black hole thermodynamics was born a little over three decades ago
from two simple questions: ``What is the most efficient way to extract
the rotational energy of a black hole?" and ``Can the second law of
thermodynamics be violated by dumping entropy
across a black hole horizon?" The answer to the first question
is that the most efficient energy extraction occurs when the horizon area
remains constant, and that moreover if the area increases the
process is classically irreversible\cite{Christo,PenroseFloyd}.
More generally, Hawking's area theorem\cite{Hawkarea} showed that the horizon
area cannot decrease provided that only positive energy can flow across
the horizon. 

Jacob Bekenstein's  proposed
answer\cite{Bek} to the second question was that it is indeed {\it not} possible
to violate the second law, provided one attributes to the horizon
itself an entropy proportional to its area. The analogy between horizon area
and entropy was already motivated by the energy extraction results and the
area theorem.  Bekenstein added information theoretic arguments,
interpreting the horizon area as a measure of the missing information
for an outside observer, and deducing that the entropy must be a constant of
order unity times the area in Planck units. He thus introduced
the generalized second law (GSL) of thermodynamics, which 
states that the sum of the black hole entropy plus the matter entropy outside
the black hole ``never decreases''\cite{Bek2}. 

The GSL took on a more precise form after
Hawking's discovery\cite{HawkThermo} 
that a black hole of surface gravity $\k$ radiates
at a temperature $T_H=\k/2\pi$.\footnote{
We use in this paper Planck units, with $\hbar=c=G=k=1$.}
When taken together with the zeroth law\cite{BCH}
which states that the surface gravity is constant 
over the horizon, and the
first law\cite{BCH} for isolated black holes, 
\beq
\label{naiveflaw}
\D M =  \frac{\kappa}{8 \pi }{\Delta A}+\Omega\Delta J\, 
\eeq
(where $M$ and $J$ are the mass and angular momentum of the black hole and 
$\O$ is the angular velocity),
this yielded the coefficient
(1/4) of proportionality between horizon area and entropy
(as well as attributing a truly thermal character to a black hole).
The GSL is then the statement that no 
physical process can lead to a decrease of the sum of the entropy 
outside a horizon $S_{outside}$ and the Bekenstein-Hawking 
entropy $S_{\rm BH}=A/4$ of the event horizon,
\beq
\label{gsl}
\Delta(S_{outside}+A/4)\ge0 \, . 
\eeq

The GSL was initially postulated with a positive energy condition in mind, 
in which case the horizon area of a black hole can only increase. 
However, it turns out that
the GSL holds even during quantum evaporation 
of the black hole via Hawking 
radiation\cite{Bek3,Hawking2},
when a negative energy flux across the horizon
produces a decrease of area.
In that case, the reason for the validity of Eq. (\ref{gsl}) 
is that Hawking radiation carries enough entropy so that the
increase of $S_{outside}$ 
outweighs
the decrease of the black hole entropy. 
Convincing arguments have been given which 
establish more generally that, in any quasi-stationary 
process, the GSL will 
hold provided the ordinary second law 
holds for matter\cite{ZurekThorne,UnruhWald,FrolovPage}.
Taken together, these results imply that black holes 
are thermodynamic systems.

The generalization of thermodynamics from black hole horizons
to de Sitter horizons was initiated by Gibbons and Hawking\cite{GH,GH2} 
shortly after the discovery of the Hawking effect,
and is by now quite well accepted, despite the observer-dependence
of de Sitter horizons. 
Nevertheless 
the corresponding generalization to acceleration horizons is 
less well accepted, 
especially with regard to the entropy they carry,
although they bear a close resemblance to 
de Sitter horizons. 
Bekenstein's
original information theoretic concept of the nature of 
black hole entropy certainly applies to all horizons,
even highly non-stationary ones,
as does the area increase theorem (see below).
The ultimate significance of black hole thermodynamics
hangs on this issue of generality, and   
we thus consider it an ideal topic for our contribution to this festschrift for Jacob.

In this article we take a broad look, both historical and fresh,
at the question how general are the laws of horizon thermodynamics.
One of the key questions is whether the local notion of horizon entropy
{\it density} is a valid concept, or whether something essentially
global is involved. 
At the classical level this question concerns the interplay of horizon
area and the gravitational field equations. At the quantum level it
concerns the role of horizon area in determining the density of states
factor in  partition functions, in 
irreversible processes,
and in transition rates.
The latter statistical role is the more 
fundamental, since it is related to the underlying quantum states, and
it gives rise to the thermodynamic laws.

Using both the classical and quantum viewpoints, 
we shall argue that local entropy density of a horizon is indeed a valid
concept, and that consequently the laws  of black hole thermodynamics 
extend quite generally to any causal horizon. 
At the statistical level a causal horizon should thus be conceived as 
an ensemble whose {\it mean} properties are the well-known geometrical ones.
This conception was pioneered by Bekenstein, who emphasized
from the outset the physical relevance of 
{\it random} exchanges which might some times lead to a decrease of the area\cite{Bek-1}. 
From this vantage point, the GSL emerges as ``a statistical law which
becomes overwhelmingly probable in the limit of a macroscopic system''\cite{Bek2}.

After all this, do we understand horizon entropy?\cite{BekMG7}
We close with a discussion of the question what does horizon 
entropy count, a question whose answer we believe,  despite some 
glimpses, remains shrouded in mystery.

\section{\large CAUSAL HORIZONS}

\begin{figure}[htb]
\centerline{\includegraphics[width=3.5in]{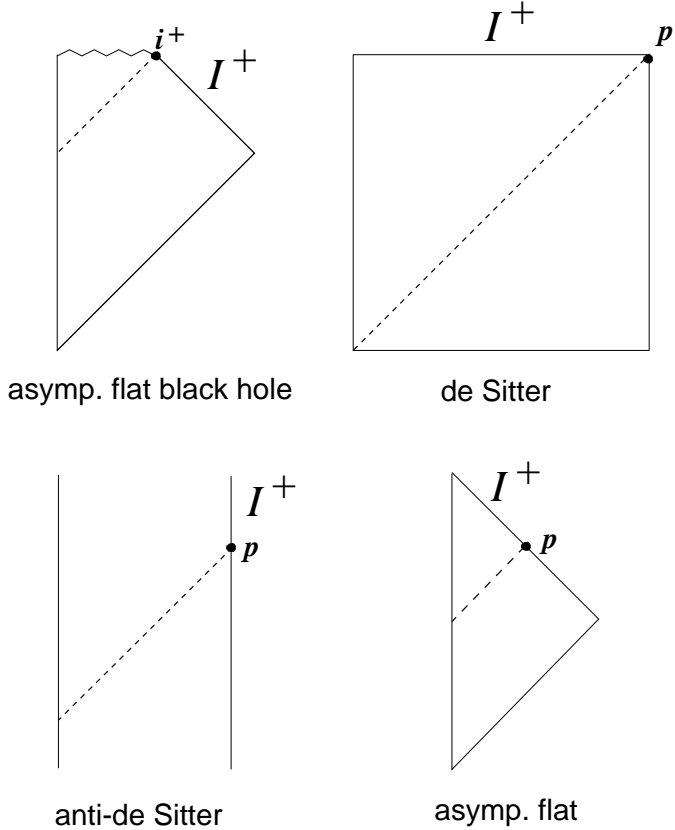}}
\caption{\small
Carter-Penrose 
diagrams of causal horizons in 
(conformal compactifications of) various 
spherically symmetric spacetimes.
Each generic point in one of the diagrams represents a 2-sphere of symmetry,
and the radial null lines make  45 degree angles with the vertical.
The points on the left hand edge of each diagram lie at 
the origin of spherical symmetry, hence they represent just a single point.
The points on the right hand edge of the de Sitter diagram lie at the
antipode of the 3-sphere, hence they also represent a single point. 
Future null infinity is denoted by \scri.}
\label{fig:ch}
\end{figure}

By a causal horizon we mean
here  the boundary of the 
past of any timelike curve $\lambda$ of infinite proper 
length in the future direction. 
This definition was 
introduced by Gibbons and Hawking\cite{GH} under 
the name ``event horizon", however we shall call it a 
``causal horizon"
to emphasize its generality, since today ``event horizon" 
refers almost exclusively to the case of a black hole event 
horizon.  
More generally, the boundary of the past of any
set of events is a kind of causal horizon. This
more local notion has not been used as much 
in the context of horizon thermodynamics
as that associated with an infinite future.
Nevertheless it does appear to have some thermodynamic
or statistical role to play, as evidenced for example in 
\cite{ees,Rovellidiamond}. 
Four examples are given in Fig. \ref{fig:ch}. 

For a black hole in asymptotically flat 
spacetime, the event horizon can be defined as the boundary 
of the past of all of \scri, however
an equivalent definition is the boundary of the past of 
{\it any} any
timelike curve that goes to future  timelike infinity $i^+$.
Equivalently, the event horizon is the boundary of the past 
of $i^+$ itself. Therefore, the black hole horizon is
defined with reference to the intrinsic asymptotic structure of the 
spacetime, without referring to a particular class of
observers.

An asymptotic de Sitter horizon is 
defined in a spacetime that is asymptotically de Sitter 
in the future. It is defined
by the boundary of the past of a  timelike 
worldline reaching a point $p$ at
\scri (which is spacelike in asymptotically de Sitter spacetime), or 
equivalently the boundary of the past of $p$.
Since there are many different points at \scri
there are many inequivalent asymptotic de Sitter horizons.
An asymptotic de Sitter horizon is thus said to be ``observer-dependent", 
in contrast to a black hole event horizon.

Our other two examples are observer-dependent in the same sense. 
An asymptotic  anti-de Sitter 
horizon, defined in a spacetime that is asymptotically anti-de Sitter 
at spatial infinity, is the boundary of the past of a timelike worldline 
reaching a point $p$ at
\scri (which is here timelike,) or equivalently the boundary 
of the past of $p$. 
An  asymptotic Rindler horizon (ARH), 
defined in a spacetime that is asymptotically flat, 
is the boundary of the past 
of an accelerated worldline that goes to a point $p$ on \scri 
(which is here null,) or equivalently the boundary 
of the past of $p$. 

Let us bring out the nature of an ARH in more detail.
The boundary of the past of 
$p$ is a cone in compactified Minkowski spacetime, however it is
actually a plane in Minkowski space. To show this
we use coordinates $(u,v,x,y)$ for which the 
line element takes the form $ds^2 = dudv - dx^2 -dy^2$.
The point $p$ could be the endpoint 
$v\rightarrow\infty$ of the light ray
$(u_0,v,x_0,y_0)$. The points $(u,v,x,y)$
to the past of $(u_0,v_0,x_0,y_0)$ are those with
$u<u_0$ and $(x_0-x)^2+(y_0-y)^2< (u_0-u)(v_0-v)$. In the 
limit $v_0\rightarrow\infty$ this includes all
points with $u<u_0$. 
The ARH is the boundary of this region, 
which is the null plane 
$u=u_0$. Thus in Minkowski spacetime an ARH is just 
what is usually meant by a ``Rindler horizon".
Note that the location of this horizon is independent of the 
transverse coordinates $x_0$ and $y_0$ of the light ray.
It is nevertheless observer dependent since it depends on $u_0$.
Note also that the horizon has infinite cross sectional area. 
An
asymptotically flat spacetime approaches Minkowski space
near \scri,  so an ARH  in general will asymptotically approach 
a Rindler horizon. 
  
A causal horizon whose generators coincide with some Killing flow
is a {\it Killing horizon}, i.e. a null hypersurface generated
by  a Killing flow. A {\it bifurcate Killing horizon} (in four dimensions)
is a pair of Killing horizons which intersect in a particular two dimensional 
spacelike cross section---called the {\it bifurcation surface}---on which 
the Killing vector vanishes. Examples of these occur in Minkowski, de Sitter, 
Anti de Sitter, and Schwarzschild spacetimes.
In the Minkowski case the pair of planes $u=0$ and $v=0$ 
(in the above mentioned coordinates) comprise a bifurcate Killing horizon 
for which the Killing field is the hyperbolic rotation generator
$-u\partial_u + v\partial_v$,
and the bifurcation surface is the $x$-$y$ plane at $u=v=0$.
This is the bifurcate Rindler or acceleration horizon, and 
the Killing field generating it is called the {\it boost Killing field}.
The ``other sheet"
of the Schwarzschild horizon is called the {\it past horizon} since it is defined
by the boundary of the future of past null infinity (in the same asymptotic region 
whose future defines the future horizon). Similarly in Minkowski spacetime for example
the other sheet is the {\it  past acceleration horizon}. 
It should be emphasized that the various examples  of Killing horizons
are all indistinguishable in a neighborhood of the bifurcation surface smaller 
than the curvature scale, hence the acceleration horizon provides a universal template 
for all of them. 

While the notion of bifurcate Killing horizon is a global one, it can be 
localized in two ways. 
Obviously, one can focus on a neighborhood of the 
bifurcation surface or even just part of it. Somewhat less obviously, 
a neighborhood of a piece of a single Killing horizon can 
be extended to a neighborhood of a bifurcate Killing horizon including 
the bifurcation surface, provided the surface gravity is constant
and nonvanishing on the horizon\cite{RaczWald1}. 
The constancy of the surface gravity---the zeroth law of horizon 
thermodynamics---can be derived either using the Einstein equation 
and the dominant energy condition\cite{BCH} or from the assumption
that a neighborhood of the horizon is 
static or
stationary-axisymmetric
with a $t$-$\phi$ reflection isometry\cite{Carter1,RaczWald2}.

In \cite{GH} Gibbons and Hawking argue that in empty 
spacetimes satisfying the Einstein equations with a cosmological constant
(or with electromagnetic fields)
stationary causal horizons are necessarily stationary axisymmetric 
Killing horizons. Their argument follows the same lines as those 
that had been given previously
for the special case of asymptotically flat black hole horizons.
(Although not stated explicitly in \cite{GH}, it seems the horizon cross sections are 
assumed to be compact in these arguments.)
They also point out that the proof in Ref. \cite{BCH} of the zeroth law for such 
Killing horizons extends immediately to the case of a cosmological constant.
Alternatively, the zeroth law follows without the use of field equations or 
energy conditions as discussed above.

\section{\large THE SECOND LAW OF CAUSAL HORIZONS}

Gibbons and Hawking\cite{GH} quote a ``second law of event 
horizons"
(whose proof is referred to a forthcoming publication that 
seems never to have appeared) which states that the area of
causal horizons cannot decrease provided that (i) what the
observer can see at late times can be predicted from a spacelike
surface (a cosmic censorship assumption), and (ii) the 
energy-momentum tensor
satisfies the strong energy condition. 
Similar theorems were quoted by Penrose\cite{Penrose} 
under both the strong and null energy conditions, and with several
different formulations of the cosmic censorship assumption
(though we have not been able to find their actual proofs in print.)
It should be noticed that all these theorems concern the 
classical version of the second law in which only positive
energy fluxes are considered. 
 
A simple but not entirely satisfactory proof  that causal horizons 
satisfy the classical area theorem
can be given under the assumptions that (i) the 
{\it null}  energy
condition holds, and (ii) the null geodesics generating the horizon 
are future complete, i.e. have infinite affine length in the future. 
This proof, which is just 
Hawking's original 
proof of the black hole area theorem\cite{Hawkarea}, 
works by contradiction. 
Suppose the congruence of horizon 
generators were converging at some point $p$. 
Then the null energy condition implies that this focusing cannot
be reversed, and will therefore produce a caustic that  
would be reached at finite affine parameter along the generator 
to the future of $p$. 
Such a caustic  is in essence an intersection of infinitesimally separated
null horizon generators, which is incompatible with the defining property of the
horizon as the boundary of the past of an infinite timelike curve.
Hence the expansion of the 
null congruence of horizon generators must be 
greater than or equal to zero everywhere. That is, the area of any bundle of 
horizon generators is non-decreasing. 

What is not entirely satisfactory about the above proof is the 
assumption that the horizon generators are future complete.
In the black hole case this assumption 
can be replaced by the 
cosmic censorship 
assumption that no singularities (to the future of a partial Cauchy
surface) are visible from \scri\cite{HawkingEllis,WaldGR}.
(To exploit this assumption it is necessary to use the device
of deforming the horizon slightly outward to another
surface containing points that are causally connected to \scri.)
For a general causal horizon similar proofs were long ago
reported\cite{GH,Penrose} (but not published) 
under different formulations of the cosmic censorship
assumption. Recently, the area theorem for general
``future horizons" has been reexamined and proved
by Chru\'sciel {\it et al.}\cite{CDGH}, 
without assuming piecewise smoothness of the horizon.

Besides such general results, 
area theorems for various special cases of horizons other than 
standard black hole horizons have been established by 
several authors. A theorem was 
proved by Davies\cite{Davies45} for the case of asymptotically de Sitter
causal horizons in Friedmann cosmologies with a cosmological
constant, under the 
assumption that the fluid satisfies the dominant energy 
condition. An area increase theorem for black holes in 
asymptotically de Sitter spacetime was proved by 
Hayward {\it et al.} \cite{Hayward} and 
Shiromizu {\it et al.}\cite{Shiromizu}.
An explicit proof of the area
increase theorem for the cosmological event horizon 
(defined as the past Cauchy horizon of \scri) 
in asymptotically de Sitter spacetimes containing a black hole
was given by Maeda {\it et al.}\cite{tatsu}.

\section{\large THE FIRST LAW OF CAUSAL HORIZONS}

In this section we consider several versions of the first
law of black hole thermodynamics and discuss how they
extend to all causal horizons. We close the section
with the argument that the ultra-local version of this first
law implies, in a precise sense, the Einstein equation
as a corresponding equation of state. 

\subsection*{Stationary comparison version}
Having shown that  causal horizons in stationary spacetimes
satisfy the zeroth and second laws, Gibbons and Hawking went on to 
derive a version of the first law. In particular, for a 
{\it pair} of infinitesimally close spacetimes 
with both cosmological and black hole horizons, 
minus the variation of the integral of 
Killing energy density  on a spatial surface bounded by the 
horizons is given by 
\beq
-\d E_{Killing} = (\k_C/8\pi)\d A_C + (\k_H/8\pi)\d A_H
+\O_H\d J_H \, .
\label{GHflaw}
\eeq 
The Killing vector $\xi^a$ employed here is 
the one that generates the cosmological horizon. 
The Killing energy density is defined by 
$T_{ab}\xi^a n^b$, where $n^b$ is the unit normal to the spatial surface, 
$\k_C$, $\k_H$, $A_C$ and $A_H$ are the surface gravities and areas 
of the cosmological and black hole horizons, $J_H$ is the
angular momentum of the black hole (defined by the integral 
of the curl of the rotational Killing field over the horizon), and $\O_H$  is
the angular velocity of the black hole relative to the cosmological horizon.
No angular momentum term appears for the cosmological horizon since
the Killing vector $\xi^a$ generates that horizon. Note that although
this Killing field has no natural normalization, that presents no difficulty 
since each term in Eq. (\ref{GHflaw}) scales with the 
undetermined normalization in the same manner. 

This version of the first law is a generalization of Eq. (\ref{naiveflaw}),
which arises when taking the asymptotically flat space-time 
limit $\kappa_C \to 0$. The generalization involves not the variation
of the ADM mass---which is not defined in a de Sitter background---but
instead the variation of the matter Killing energy. It can thus be used 
to describe variations in the asymptotically flat case between
two stationary configurations at {\it fixed} ADM
energy, in which the matter energy is redistributed, with some
going into the black hole. Formulated this way, the first law is 
more local, since no reference to what is happening at infinity
is required. 

\subsection*{Physical process version}
Both of the previously mentioned versions of the first law
Eq. (\ref{naiveflaw}) and Eq. (\ref{GHflaw}) involve comparisons
of two stationary solutions to the field equations.
A yet more local version of the first law 
emerges when one enquires into the dynamical
process by which the horizon area adjusts to a 
small flow of energy across the horizon. 
This relation, which 
Wald\cite{Waldphysproc} called the 
``physical process version"  of the first law in the case of 
black hole horizons, was first established in \cite{HawkHart}
(see also \cite{Carter2} for a comprehensive review.)
The physical process version of the first law states that
\beq
\label{ppflaw}
\d S=\d E_H/T_H,
\eeq
where $\d S=\d A/4$ is one fourth the horizon area change, 
$\d E_H=\int_H T_{ab}\xi^a d\S^b$ 
is the flux of  ``energy" 
across the horizon, defined 
with respect 
to the 
horizon generating
Killing field $\xi^a$ of the 
stationary 
background
spacetime
which is being perturbed,
and $T_H$ is $1/2\pi$ times the surface gravity 
$\k=|\nabla_a \xi|$ of 
the horizon generating 
Killing field.  

To exhibit its local nature, let us review the
derivation of the physical process first law.
We first discuss the approach using affine
parametrization and then discuss the 
approach using Killing parametrization.
The expansion $\theta$ of the null congruence of
horizon generators measures the fractional rate
of increase of the cross-sectional area element with
respect to an
affine parameter $\l$.
Hence the change in the area of a bundle of generators
over a finite range of affine parameter is
\beq
\D A = \int_B \theta\,  d^2\!\!A\, d\l,
\label{DA}
\eeq
where the integral is over a patch $B$ of the horizon
consisting of the finite range of the chosen generators.
This integral can be related to the energy flux 
using the identity $\theta= d(\l\theta)/d\l - \l d\theta/d\l$,
together with the Raychaudhuri equation for 
$d\theta/d\l$ and the Einstein equation.
The Raychaudhuri equation is
\beq
\frac{d\theta}{d\l}= -\frac{1}{2}\theta^2 -\frac{1}{2}\sigma^2 -R_{ab}k^ak^b,
\label{Ray}
\eeq
where $\sigma^2=\s_{ab}\s^{ab}$ is the squared 
shear of the congruence and $k^a$ is the
affine tangent vector to the congruence,
and the Einstein equation reads
\beq
R_{ab}-\frac{1}{2}Rg_{ab}=8\pi  T_{ab}.
\eeq
Eq. (\ref{DA}) thus yields
\beq
\D A=[\l\theta A]_1^2 +  \int_B \frac{1}{2}(\theta^2 +\sigma^2)\,  d^2A\, d\l,
+\int_B 8\pi T_{ab} \l k^a d\S^b.
\label{DA2}
\eeq

Eq. (\ref{DA2}) is an identity satisfied by any null congruence in a spacetime
satisfying the Einstein equation. It becomes the first law when we specialize
to a situation involving a small change of a stationary background spacetime,
with the congruence corresponding to a small deformation of the Killing horizon.
We now see how this happens by examining each term in the equation.

If the small deformation is caused by some matter stress tensor at order $\epsilon$,
then the metric perturbation and therefore $\theta$ and $\sigma$ will be of 
order $\epsilon$. Hence the term $(\theta^2 +\sigma^2)$ can be 
neglected.\footnote{Actually, neglect of this term is not justified if
the generators develop caustics or if the integration region extends
to infinity, as in the case of an asymptotic Rindler horizon, 
as will be discussed below.}

If the upper limit of integration is chosen in a region where the horizon is again
stationary, then 
the net area change of a given bundle of generators must be finite, 
hence $\theta$ must vanish faster than $1/\l$, so the term  $(\l\theta A)_2$
vanishes. To understand the behavior of the lower limit it is helpful
to relate the affine parameter $\l$ to the Killing parameter $v$ of the 
background stationary spacetime. 
In general the relation between affine and Killing parameters
on a Killing horizon is 
\beq
\l=ae^{\k v}+b,
\label{lv}
\eeq
where $a$ and $b$ are arbitrary constants.
If we exploit the shift freedom of the affine parameter 
to set $b=0$, then as the Killing
parameter $v$ goes into the past, the affine parameter $\l$ 
goes to zero exponentially fast. The term  $(\l\theta A)_1$
is already first order in $\e$ from $\theta$, hence to first
order we can use this relation between the background 
values of $\l$ and $v$ to conclude that this term is 
exponentially suppressed provided the stretch of (background) 
Killing time over which the
first law is being applied is long compared with $\k^{-1}$. 
This temporal restriction of the applicability of the first law
corresponds to an ``equilibration time" (albeit backwards in
time). Indeed, the expansion defined with respect to Killing time
also vanishes exponentially.

The last step is to relate $\l k^a$ in the energy flux term 
to the Killing vector $\xi^a$. Since the energy is a first order perturbation,
this relation can be evaluated in the strictly stationary background on the 
Killing horizon.  We thus have $\l k^a = \l(dv/d\l)\xi^a$, where $v$ is the Killing
parameter. From the relation (\ref{lv}) with $b=0$ we have $\l (dv/d\l) =  \k^{-1}$, 
hence
\beq
\l k^a = \k^{-1}\xi^a.
\label{k-xi}
\eeq
When substituted into Eq. (\ref{DA2}) 
this yields the physical process form of the first law Eq. (\ref{ppflaw})
when the patch $B$ is taken to be a piece of the horizon bounded
in the past and the future by two complete cross-sections.

\subsection*{Adiabatic version}
In the preceding formulation of the first law for physical processes
it was assumed only that the neighborhood of the horizon 
is nearly stationary, so that the perturbation is {\it small}. 
If one assumes further that it is {\it adiabatic}, then an even 
more local form of the first law applies.
As pointed out above, the horizon equilibration rate is given by
$\kappa$, hence adiabatic means slow with respect to that scale.
To extract the adiabatic limit, it is useful to re-express
the Raychaudhuri equation (\ref{Ray}) 
in terms of derivatives with
respect to the Killing parameter $v$ rather than the affine parameter.
This produces the equation
\beq
\frac{d\hat{\theta}}{dv}= \kappa\hat{\theta}-\frac{1}{2}\hat{\theta}^2 -
\frac{1}{2}\hat{\sigma}^2 -R_{ab}\hat{k}^a\hat{k}^b,
\label{Ray2}
\eeq
where $\hat{\theta}=(d\l/dv)\theta$, $\hat{\sigma}=(d\l/dv)\sigma$, 
$\hat{k}^a=(d\l/dv)k^a$ and 
$\kappa = (d^2\l/dv^2)/(d\l/dv)$ (cf. Eq. (\ref{lv})). 
For a slow perturbation we have  $\hat{\theta} \ll  \kappa$,  $\hat{\sigma} \ll  \kappa$,
and $d\hat{\theta}/dv \ll  \kappa\hat{\theta}$, 
so both the left hand side and the second order terms in the expansion 
and shear can be dropped, which yields 
\beq
\hat{\theta}\approx \kappa^{-1}R_{ab}\hat{k}^a\hat{k}^b.
\label{Rayslow}
\eeq
When this is substituted in Eq. (\ref{DA}) and the Einstein equation is used,
we see that,  when the evolution is adiabatic, the first law Eq. (\ref{ppflaw}) 
holds over any patch {\it and} over any stretch of time
of the slowly evolving horizon. 

\subsection*{Extension to all causal horizons}
While the preceding derivations 
apply for black hole horizons, 
they are localized
on the horizon and  make no reference to spatial infinity.
Therefore it is tempting to apply them to all causal horizons.
Possible obstructions concern the normalization of the Killing
vector, the possibly infinite horizon area, and convergence of the upper limit.
Let us consider these in turn.

First, 
the energy flux $\d E_H$ 
and the temperature $T_H$ 
scale in the same way under a change of normalization of the 
Killing field. Hence, as in Eq. (\ref{GHflaw}),
this normalization cancels out so no choice is required.
(In the stationary black hole case the horizon generating Killing field
is a linear combination of a time translation and a rotation,
$\xi^a = (\partial/\partial t)^a +\O_H(\partial/\partial \phi)^a$.
The time translation Killing field $(\partial/\partial t)^a$ is normalized at infinity,
so the change of  ``energy" in Eq. (\ref{ppflaw}) is expressed as $\d M-\O_H \d J$.)

Second, the derivation of the first law can be applied to a bundle of generators forming
a finite subset of the horizon, so no infinite area need be confronted.
Moreover,  the change of 
total
area $\d A$ can be finite, even if the total area itself is infinite. This change
can be defined 
by considering first a finite bundle of
horizon generators whose cross-sectional area is then taken to 
infinity.\footnote{See \cite{HHR} for  
an example of computing
the finite difference of acceleration horizon areas 
when comparing two different spacetimes.}

Third, convergence of the upper limit $(\l\theta A)_2$ of Eq. (\ref{DA2})
can be analyzed in each bundle of generators. As explained above
this upper limit vanishes if and only if the
cross sectional area of any finite bundle is finite at infinity.
This corresponds to the condition that the horizon is 
in equilibrium in the asymptotic future. 
An ARH satisfies this condition, since it is just a Rindler horizon
at infinity. (We have not checked whether the horizon of a point at 
\scri in AdS is asymptotically stationary.)
\begin{quote}
{\it The physical process
first law thus applies to all situations
in which a causal horizon goes through an evolution that 
is nearly stationary and settles down in the future.}
\end{quote}

\subsection*{Asymptotic Rindler horizons}

In the limit that the cosmological constant goes to
zero, the stationary comparison version (\ref{GHflaw}) 
of the first law for cosmological horizons
becomes the first law for Rindler horizons. 
To illustrate this, we suppose a rest mass $\d m$ 
follows a Killing orbit on which the norm of the 
Killing field is $\xi$. The Killing energy of 
this mass is $\xi\, \d m$, so the first law (\ref{GHflaw})
implies that the disappearance of the mass increases
the horizon area by $\d A_C= (8\pi\xi/\k_C)\, \d m$. In de Sitter
space we have $\k_C/\xi=\sqrt{a^2+\L}$, where $a$ is the
proper acceleration of the Killing orbit. If the mass sits 
at the center where $a=0$, then in the limit as $\L$ goes to
zero the area change diverges. If instead $a\ne 0$ then the 
ratio  $\k_C/\xi$ approaches $a$ and the area change is 
finite. It is interesting to note that, when $\L$ is
nonzero,  $\k_C/2\pi\xi$ is equal to the local 
Unruh/de Sitter temperature $T_{local}$
that would be experienced
by an accelerated mass in the Euclidean vacuum state\cite{Pf}.
(As $\L$ goes to zero this becomes just the Unruh temperature
for the accelerated mass and as $a$ goes to zero it becomes
just the de Sitter temperature.)
The disappearance of the mass thus increases the horizon
entropy by an amount equal to $\d m/T_{local}$ 
for all values of $\L$ and $a$.

Unlike the stationary comparison version,
%
the physical process version (\ref{ppflaw}) of the first law 
does {\it not} hold for an asymptotic Rindler horizon 
since the `nearly stationary'
condition cannot be satisfied.
We now explain the reason
for this in two ways, first from the perspective of the ARH itself,
and then by describing the ARH as the infinite mass limit
of a Schwarzschild horizon.

Suppose a planet in free fall
wanders across the ARH. Although the boost energy
flux of the planet is finite, it can be shown that the
net area change caused by the passage of the planet is
infinite. This is essentially because every null geodesic on
the ARH---no matter how far from the planet in the transverse
direction--- is eventually focused backwards in time to a focal 
point on the line
behind the planet, opposite to the direction of the endpoint
of the ARH. That is, all the horizon
generators originate on the line behind the planet. Hence an infinite
area is added to the ARH 
so evidently the first law (\ref{ppflaw}) is not satisfied.

The reason the physical process first law is not satisfied
is that the process is not nearly stationary.
The Raychaudhuri equation (\ref{Ray})  of course applies,
but the terms neglected to arrive at the first law cannot be
neglected. For example this is clear from the fact that
there are focal points, where the expansion goes to
infinity and hence $\theta^2$ cannot be
neglected.\footnote{A similar but simpler
example of a non-nearly stationary process is the formation of a
black hole by an infalling  thin spherical shell of mass $M$. The
horizon grows in area from the focal point in the center where it
originates, and this area growth is not properly accounted for by
the first law. However, if a second such shell with mass $\D M\ll M$
follows, the transition does obey the first law.}
Moreover, since all generators originate on the line of
caustics, the ultimate area of the annulus of generators between asymptotic
cylindrical radii $\rho$ and $\rho+\D\rho$ must grow from zero without any energy
having crossed the congruence. According to (\ref{DA2}) all of this
area must come from integrating the squared expansion and shear.
The shear will dominate (since it is the shear that generates the
expansion in this case\cite{HawkHart}). 
For any bundle of generators
this produces a finite final area, however when integrated over an
infinite cross section and infinite affine parameter range it produces
an infinite total area change. Thus, even though the squared shear is
$O(\e^2)$ in the perturbation, it has an infinite net effect.

Further insight is offered by viewing the ARH as the infinite mass
limit of a Schwarzschild horizon of mass $M$. The planet crossing
the ARH can be modeled by a planet that falls in freely from infinity
and crosses the Schwarzschild horizon. This situation has been
analyzed in \cite{Membrane}, where it was shown that the process
is nearly stationary for the purposes of the first law only if
$R\gg \sqrt{mM}$, where $R$ is the radius of the planet and 
$m$ is the mass.\footnote{This condition can also be expressed by
saying that the surface area of the planet is much larger than the
increase of the horizon area according to the first law.}
In the limit of $M$ going to infinity, this requires
that the planet have an infinite radius.  Hence,
for any finite sized planet crossing, an ARH will not satisfy the
physical process version of the first law.
The fact that the area increase is {\it infinite} can also be
understood using the infinite mass limit, since the planet
comes in from infinity with finite Killing energy $m$ and
the surface gravity $\k=1/4M$ goes to zero. Hence, if
the physical process
form of the first law were to apply, the area change would be
given by $\d A\sim (8\pi/\k)m=32\pi mM$, which diverges as $M$ goes to
infinity.\footnote{As the planet falls, a small fraction
($\sim 0.01 m/M$)
of the rest mass is radiated as gravitational waves, hence the Killing
energy crossing the horizon approaches $m$ as $m/M$ goes to
zero.}
Failure of the nearly stationary condition would only add to this divergence.

It should be emphasized that although the nearly stationary 
physical process first law (\ref{ppflaw}) does not hold, the 
expression (\ref{DA2}) for the horizon area change remains valid.
This should be considered an extension of the first law to 
non-equilibrium thermodynamics, as discussed at the classical
level in \cite{HawkHart,Carter2} and at the quantum statistical level in
\cite{Candelas-SciamaPRL,Sciama}. It would be interesting to pursue 
this line of thought further.

\subsection*{Local Rindler horizons and the Einstein equation of state}
We have just seen how the Einstein equation enforces the first
law for near-stationary transitions of causal horizons. 
To conclude this section let us show how this logic
can be turned around to {\it derive} the Einstein equation
as an ``equation of state" of spacetime\cite{ees}.
To render the argument entirely local we make use of 
the notion of  a ``local Rindler horizon" (LRH). An LRH is
defined as the boundary of the past (on one side) of a spacelike
two-surface element $\cal S$ through a point $p$,
with 
$\cal S$ 
adjusted so that the generators of the LRH have vanishing
expansion and shear at $p$. Thus 
the LRH reaches equilibrium at $p$.
The area change of a small bundle of generators approaching the 
``equilibrium point" $p$ is given by Eq. (\ref{DA}), with 
\beq
\theta\approx -\l\, R_{ab}k^a k^b,
\eeq
where the affine parameter $\l$ has been set to zero at $p$,
and $k^a$ is the affinely parametrized horizon generator.
If we restrict attention to a sufficiently
small neighborhood, then the effect of any classical energy momentum 
tensor makes only a small perturbation of the local Minkowski
space, hence the near-stationary analysis holds. 
According to Eq. (\ref{k-xi}) we can replace $-\l k^a$ by $\k^{-1}\xi^a$, where 
$\xi^a$ is the boost Killing field on the LRH. 
The first law Eq. (\ref{ppflaw}) will
hold on the small bundle of generators,
with $T_H$ equal to the Unruh temperature associated with $\xi^a$,
 if and only if at $p$ we have
\beq
R_{ab}k^a k^b = 8\pi T_{ab}k^a k^b.
\eeq
If  this is to hold for all LRH's through $p$ then it must hold
for all $k^a$, which implies that 
$G_{ab}=8\pi T_{ab} + f g_{ab}$, where $f$ is an arbitrary function.
The Bianchi identity and the conservation of $T_{ab}$ then imply that
$f$ is a constant, which can be identified with an undetermined
cosmological constant.

It is difficult to resist concluding from this argument that the horizon entropy 
density proportional to area is a more primitive concept than the classical Einstein
equation, which appears as a thermodynamic consequence of the 
interplay of entropy and causality. From the point of view of the
quantum action principle this is conclusion is not so surprising. The action 
gives rise to the entropy at the quantum statistical level, 
and to the field equations at the level of the classical mean field.
It is to the former, deeper level of description that we now turn.

\section{{\large STATISTICAL MECHANICS OF CAUSAL HORIZONS}}

In this section we discuss, at the quantum statistical level,
the extension of the notion
of black hole Boltzmann entropy to 
stationary states of all causal horizons, i.e. to Killing horizons.
The starting point for this generalization is 
the work of Gibbons and Hawking\cite{GH2} which
showed  that semiclassical evaluation of the partition function
for the gravitational field yields $A/4$ for the entropy of both black
holes and de Sitter space. Here we examine the
role of horizon entropy in determining the rates for 
dynamic processes,
and we argue that this role extends in particular
to acceleration horizons. We also briefly 
consider a few approaches to explicit ``state counting" 
of horizon degrees of freedom.

\subsection*{Pair creation}
A first indication in favor of the 
 extension of the statistical notion of black hole entropy 
to all causal horizons arises from 
the pair creation probability for magnetically 
charged black holes in a magnetic field\cite{Garfinkle}.
This probability is given as a product of two exponentials:
\beq
\label{bhpc}
P(q,M ; B) \propto e^{A/4} \times  e^{\D A_{accel}/4}\, .
\eeq
The second 
factor 
contains $\D A_{accel}(q, M, B)$:
the change of area of the acceleration horizon\cite{HHR}
which is associated with the creation of the black hole pair.
This term persists even when no black holes but only
monopoles are created, and it reproduces the well known instanton
contribution $\exp(-\pi M^2/ q B)$ in the test field approximation
$G \to 0$ where $G$ is Newton's constant.
The first factor tells us that the probability is enhanced\cite{Garfinkle,HHR} 
over that of monopoles
by $\exp(A/4)$, where $A(q, M)$ is the horizon area of 
a black hole of mass $M$ and magnetic 
charge $q$.\footnote{One might have thought that it should
be the total area of the {\it two} black holes appearing here, 
however that is not the case. 
The reason is that the pair carries the quantum numbers of 
the vacuum, so their quantum degrees of freedom are strictly 
correlated\cite{Mapa97}. 
Thus the degeneracy of states is that of a single black hole.}

This is consistent with the usual statistical interpretation of $A/4$ as the
logarithm of the number of black hole states.
From Eq. (\ref{bhpc}) 
we see that the two area changes determine 
the probability in the same manner. 
It is therefore compelling to apply the same statistical interpretation
to both of them.
Perhaps the strongest indication in favour of treating
both areas on the same footing arises from the fact that an accelerated
black hole will be in a {\it statistical ensemble}
since it will 
experience Unruh radiation. The random exchanges of 
thermal photons 
between the
acceleration horizon and the hole will bring the whole system into
equilibrium\cite{Mapa97}.
It is to be noticed that the equilibrium situation is determined by 
Eq. (\ref{bhpc}) and corresponds to the maximum probability
where $\partial_M (A + \Delta A_{acc}) = 0$,
i.e. when Hawking and Unruh temperatures agree. 
In this $S_{system} = (A + \Delta A_{acc})/4$ does act as the entropy of the whole system.
The random energy exchanges will also {\it spread} both the black hole mass 
and the area of the acceleration horizon around their mean values.
The spread in the mass of the hole is related to the heat capacity and it 
is given, 
as usual, by  
$(\partial_M^2  S_{system})_{equil}^{-1/2}$.
We emphasize that the conclusion that the 
acceleration horizon is in an ensemble
requires no assumption other than that 
the creation probability Eq. (\ref{bhpc}) also delivers 
the equilibrium distribution. 
The considerations of the following subsection 
support this natural assumption.
 
Consideration of pair creation of black holes in de Sitter 
space yields again a probability 
given as the exponential of two area changes\cite{desitpc}. 
In this case, 
people would 
likely
agree 
that both terms possess a statistical interpretation. 
Despite the similarity between the two cases,
Hawking and Horowitz\cite{HH}
nevertheless assert that 
the acceleration horizon area
should not be interpreted as counting 
an entropy.
They give two reasons. The first is that the
usual evaluation of the entropy 
from  the partition function 
cannot be performed since
the period of the Euclidean section is
fixed and cannot be
varied. The second is that the 
acceleration horizon is observer dependent
and hence the information behind the horizon
``can be recovered by observers who simply stop accelerating".
Restated, they assert the unitarity of the evolution
on a foliation cutting across the acceleration horizon.
The last two statements can be made
for a black hole horizon (or a de Sitter horizon) however, 
so they do not seem to preclude the association of area with
entropy for the acceleration horizon. 

\subsection*{Transition rates for systems in contact with a Killing horizon}

Further evidence for the statistical entropy of causal horizons is obtained 
from \cite{MaPa} where it is explained why the transition rates for systems
in contact with a Killing horizon are governed by changes in the horizon area. 
The analysis
applies to black hole horizons with the
evaporation process itself, to acceleration horizons
with the Unruh effect experienced by accelerated systems,
and to cosmological horizons with the processes related to the de Sitter 
temperature.
In all cases, when taking into account the 
gravitational back-reaction, 
the excitation rates are governed by 
$\exp(\d A/4)$ 
in place of 
the thermal Boltzmann factor
$\exp(-\d E_K/T_H)$
which is
known to govern Hawking radiation\cite{HartleH} for example.
In these expressions,
$\d E_K$ is the change in Killing energy, $T_H$ the horizon temperature 
measured in the same units as $\d E_K$, and
$\d A$ is the corresponding area change. 
In the test field approximation,
$\d A$ is related to the matter change by Eq. (\ref{GHflaw}). 
Hence 
we recover the transition rate in a thermal state,
as 
previously obtained in the case of Hawking radiation
or transitions of an accelerated detector\cite{Unruh76}.
The novelty consists of taking into account the back-reaction of 
the spacetime geometry in the evaluation of individual transition amplitudes (like
one does it with recoil effects\cite{recoils}), and no longer in the mean
as one does when using the semi-classical Einstein equations. 
As a byproduct, one notices that
canonical distributions (governed by Hawking or Unruh temperature) 
are 
replaced by microcanonical distributions
in which the heat capacity of the horizon is 
properly accounted for.

To understand how this replacement  occurs, 
recall that
the action 
must sometimes be
supplemented by boundary terms in order to have
a well-defined action principle,  
since
the action 
should be stationary when varied within the class of metrics
under consideration. 
Consider $S_h$, the 
action for 
matter+gravity in its hamiltonian form and applied to the 
class of metrics which belong to the ``one black hole sector''\cite{Carlip-T},
namely metrics which are asymptotically flat and which also 
possess an inner horizon. The total variation of $S_h$ reads
\beq
\d S_h = \int\! d\Sigma_3(\pi \d g + p \d q)|^2_1 + {\rm bulk \, terms} 
+ \d M (t_2 - t_1) - { \d A\over 8 \pi}
(\Theta_2 - \Theta_1)\, . 
\label{var}
\eeq
The first terms give the initial and final momenta of gravity and matter respectively.
The bulk terms vanish on-shell. The third term arises from the outer boundary. 
It is given by the variation of the ADM mass times the 
lapse of asymptotic proper time. The last term arises from the  inner 
boundary and it is given by the change of the horizon area times the lapse
of the hyperbolic angle $\Theta$. (On-shell
and for stationary metrics,
 $d \Theta/dt = \kappa$ gives the surface gravity.) 
For
the class of processes involving 
exchanges between a black hole and surrounding matter at fixed 
ADM energy $M$,
the outer boundary term vanishes. Hence one should {\it not} add to $S_h$ the
energy term $-\!\! \int\! dt M$. On the 
other hand, 
since 
the area 
then adjusts itself
dynamically, it should not be fixed. Thus, one must get rid of the
inner boundary term in Eq. (\ref{var}) 
by working with 
the 
(Legendre transformed)
{\it exchange} action
\beq
S_{exchange} = S_h + { A \over 8 \pi } \Theta \, .
\eeq

As noticed by Carlip and Teitelboim \cite{Carlip-T}, 
quantization of this action yields for the matter + gravity wave function
the unusual Schr\"odinger equation 
\beq
i  {d \over d \Theta} \Psi =  - { \hat A \over 8 \pi } \Psi \, ,
\label{Schrodtheta}
\eeq
where $\hat A$ is the operator valued horizon area which acts on both
the matter and gravity sectors. 
The system at fixed ADM energy thus 
evolves according to $\Theta$ and not the 
time at infinity. 
Eq.  (\ref{Schrodtheta}) is remarkable.  
It shows that the area itself is the operator-valued geometrical quantity
that replaces the Hamiltonian in governing exchanges.  

Two applications of  Eq.  (\ref{Schrodtheta}) were
considered in \cite{MaPa}.
First is the thermalisation of a two level atom
held at some fixed distance from a black hole. This 
type of situation  
also applies to
the thermalisation of inertial systems in de Sitter space and leads to 
expressions that
can be put in parallel to the static version of the first law
Eq. (\ref{GHflaw}). The other case concerns self-gravitating effects within Hawking radiation
as studied in \cite{KW,KVK}. 
The results found there emerge simply and transparently
when the system is analyzed using the conjugate
variables $\Theta$ and $A$ rather than the pair of asymptotic 
variables $t$ and $M$.

The essential new point is that, via the gravitational back reaction, 
each matter energy eigenstate $j$ is correlated
to an area eigenstate with eigenvalue $A_j$. Thus it 
evolves 
according to (\ref{Schrodtheta})
with a 
$\Theta$-time dependence given by $\exp(i A_j \Theta/ 8 \pi)$. 
(The entanglement of the 
gravitational 
subspace of eigenvalue $A_j$
to the matter state $j$
is not specific to the present case: because of the constraints,
this gravity-matter entanglement always occurs when working 
with the solutions of the Wheeler-DeWitt equation\cite{PTwdw}.)

As an illustration
the system can be taken to be a two-level atom held fixed
at some distance from the hole (for a more complete
discussion see \cite{MaPa}).
To mediate the interactions 
we 
introduce 
a radiation field
which allows for transitions 
from ground state $j$ to excited state $j'$
by 
absorption or emission
of a photon.
To first order in the coupling between the atom and the radiation field,
the transition amplitude from $j$ to $j'$ 
accompanied
by the emission of a photon of frequency $\Omega$ is given by
\bea
{\bf T}_{j \to j'+ \Omega} &=& C \int\!\! d\Theta \, \, \Psi_{j} \, (  \Psi_{j'} \, \, \phi_\Omega )^*
\nonumber\\
&=& \tilde C \int\!\! d\Theta \, \, e^{i(A_j - A_{j'}) \Theta/8 \pi} \, 
 \, e^{-i\Omega e^{-\Theta}}\, .
\label{newT}
\eea
In the first line, the wave function $\Psi_{j}$ characterizes the time evolution of 
the initial state of the system: a black hole plus the detector in its ground state
with the radiation field in its vacuum state. The product in parenthesis 
characterizes the final state: 
the black hole and the excited detector plus the photon of frequency $\Omega$.
Their wave functions factorize because we neglect the gravitational back-reaction of the radiation. 
(This is a legitimate assumption when computing the rates to leading order in the 
back-reaction.)
The overall constants $C$ and $\tilde C$ will play no role since we shall be interested 
in
the {ratio}
${\bf R} = |{\bf T}_{j \to j'+ \Omega} / {\bf T}_{j' \to j+ \Omega}|^2$ which determines the
{\it equilibrium} distribution. 
In the second line, we have grouped together the two `unperturbed' stationary states
with their corresponding area-eigenvalues. The last factor
is the wave function of a photon of (Kruskal) frequency $\Omega$. Its properties are 
fixed by the following considerations. Since we are interested in near equilibrium transitions,
the 
initial 
state of the radiation field must be stationary and regular on the horizon.
It must therefore be either the Hartle-Hawing vacuum\cite{HartleH}
or the Unruh vacuum, depending on the condition one imposes on 
infalling configurations. 
It is most economical 
(though physically less direct)
to characterize one-particle states 
by making use of modes which individually incorporate the requirement of regularity
at the expense of being non-stationary. (The other option would be to work with 
stationary modes (like Schwarzschild modes 
$e^{-i \lambda \Theta}$)
and to express the Unruh or Hartle-Hawking state
as a thermal state in terms of these modes.)
When working with the first option, the modes are
of the form $\exp(-i \Omega U)$, where $U$
is a null outgoing coordinate which is regular on the future 
horizon (see e.g. \cite{Primer} for a discussion of the various relevant modes).
When evaluated at the (stationary) location of the two level atom, 
one has $U(\Theta) = - c \, e^{-\Theta}$,
where the constant $c$ can be put to 
unity by a shift in $\Theta$. 

Before proceeding, it is of interest to display the corresponding expression
one would obtain by working in a fixed background. Instead of Eq. (\ref{newT})
one would have
\bea
T_{j \to j', \Omega} &=&  C' \int \!\! dt  \, \psi_j \, \, (\psi_{j'} \, \phi_\Omega)^*
\nonumber\\
 &=& \tilde C' \int \!\! dt  \, e^{-i(E_j-E_{j'})t}  \, e^{-i\Omega e^{-\kappa t}} \, ,
\eea
where $E_j$ and $E_{j'}$ are the eigen-energies of the two atom states
measured with the asymptotic time which defines the surface gravity $\kappa$.

When working with the non-stationary modes, 
the ratio  $\bf R$ of transition rates can be simply evaluated,
at fixed $\Omega$, by 
recognizing from (\ref{newT}) that ${\bf T}_{j' \to j+ \Omega}$ is related
to ${\bf T}_{j \to j'+ \Omega}^*$ by 
a shift $\Theta\rightarrow \Theta-i\pi$ of the $\Theta$ contour.
This yields
${\bf R} = \exp(A_{j'}-A_{j})/4$ 
for the case
when 
the gravity-matter entanglement has been taken into account in the amplitudes.
On the other hand when using the background fixed transitions the 
ratio of the rates is 
$R = \exp-(E_{j'}-E_{j}) (2 \pi/\kappa)$, 
which 
gives rise to
a Boltzmann distribution with the horizon temperature  $\kappa/2 \pi$. 
As mentioned above, the first law Eq. (\ref{GHflaw}) guarantees 
that 
these
results agree to first order in $G$. 
The only difference is that, in the first case, the specific heat of the horizon has been 
taken into account through the gravity-matter entanglement of the wave functions.
Let us emphasize that in both cases, the ratio of the transition rates has been 
computed unconventionally by making use of the (non-stationary) Kruskal 
modes. The stationarity of the processes
follows from the fact that the above ratios are independent of 
$\Omega$.

The principle of detailed balance implies that
the ratio $p_{j'}/p_j$ of equilibrium occupation probabilities for
the detector + horizon system, with the field
in the Unruh or Hartle-Hawking state, is given
by the ratio ${\bf R} = \exp(A_{j'}-A_{j})/4$
of transition rates.
The equilibrium
distribution is also the microcanonical
distribution, hence we infer that the number of
states in a configuration with area $A$ is 
proportional to $\exp A/4$.
Since the above analysis
is couched in terms of $A$ and $\Theta$, which
are defined at the horizon,
it is entirely local
to the neighborhood of the Killing horizon and
therefore takes the same form for black hole, de Sitter, and
acceleration horizons.
It was therefore concluded
in \cite{MaPa}
that a quantum statistical interpretation of the entropy $A/4$
should apply to all ``event horizons".

A difference between the 
Rindler horizon and  both the 
black hole and de Sitter ones is that in the 
former case the temperature of the horizon is perceived
as a physical temperature only  by 
observers who follow the 
{\it accelerated} worldlines of the 
horizon generating Killing field (i.e. it
is the Unruh temperature), while
in the black hole and de Sitter 
cases there are also unaccelerated observers who perceive a nonzero 
temperature. This difference plays no role in the 
above analysis however.

\subsection*{Other approaches to counting horizon states}

Of the many other approaches to calculating black hole entropy from 
quantum first principles,  it seems to us that none present any evidence 
that the notion of universal horizon entropy density is invalid. 
For examples we briefly consider here the approaches of 
Carlip\cite{Carlip} using 
the structure of the near-horizon component of the diffeomorphism
group, the canonical quantum gravity calculations in the loop
approach\cite{Rovelli}, and the calculations of 
string theory.

Carlip's approach, being based on representations
of the residual {\it local} diffeomorphism symmetry modulo horizon
boundary conditions, is as applicable to any Killing
horizon as it is to black hole horizons. It does not 
rely on the global properties of the horizon, but instead appears 
consistent with the notion of horizon entropy density.
The loop quantum gravity approach as currently 
formulated involves an induced
Chern-Simons theory on the horizon, and applies to both
compact black hole and de Sitter horizons.  For technical reasons
however it does not apply to non-compact Rindler horizons or even to
a Rindler horizon compactified to a 2-torus. It is not yet clear
what result would be obtained for the entropy of 
a Rindler horizon using this approach.

The string calculations are of two types. In the weakly coupled 
D-brane context, near-extremal configurations at weak coupling 
are found to have the same entropy as the black hole with the 
corresponding charges at strong coupling (for a review
see \cite{HoroChandra}).  
To apply this reasoning
to other causal horizons seems problematic, since it is
unclear how to identify an appropriate weakly coupled cousin.
The second (and related) 
string approach is in the context of the AdS/CFT
duality, wherein the black hole entropy is identified with ordinary 
thermal entropy in the CFT (for a review see \cite{Magoo}). 
Hawking, Maldacena, and Strominger\cite{HMS} 
applied these ideas to understand the entropy in
a two dimensional de Sitter brane world, embedded in 
AdS, in terms of entanglement entropy of the conformal field
theory vacuum across the de Sitter horizon, and they made arguments
for the extension of this result to higher dimensions.
Somewhat similarly, Das and Zelnikov\cite{Zelnikov} 
have identified the mapping between the Unruh radiation for an accelerated
observer in AdS and a thermal appearance of the 
ground state in the CFT for a corresponding class
of observers. Presumably the acceleration horizon entropy can be identified 
with some definite CFT entropy along these lines, although that 
has not been shown explicitly.  

In summary, it is fair to say that, wherever the question can be 
addressed, all of these approaches to counting black hole states
apply in principle equally well to all Killing horizons.
This seems to be consistent with the universal interpretation of 
horizon entropy we are advocating here. 

\section{\large WHAT DOES HORIZON ENTROPY COUNT?}

We have argued that the entire framework of 
black hole 
thermodynamics and in particular the notion
of black hole entropy extends to any causal 
horizon. What are the 
implications of this conclusion?

First of all, our attention is deflected away from the black 
holes and towards 
the horizons in black hole thermodynamics. It is sometimes 
considered a mystery 
how a black hole horizon could be capable of carrying so 
much entropy when after 
all it has no local significance in but is rather defined 
teleologically in 
terms of the future evolution of the spacetime. For example, 
it is regarded as 
puzzling that when a star collapses and forms a black hole, 
the entropy suddenly 
rockets up to a value many orders of magnitude greater than 
it was in the star, 
``just because" the horizon has formed. 
This becomes much 
less mysterious when 
it is realized
that in essence the black 
hole really has 
nothing to do with it. Any causal horizon is endowed with a 
surface entropy density of 1/4. 

The realization that horizon entropy is an
intrinsically observer dependent notion
raises the obvious question
of what are the states that the horizon entropy counts?
The notion that it counts the number of 
{\it internal} configuations, i.e. configurations
{\it behind} the horizon, was argued against
in \cite{nature} on various grounds.
It seems only even possibly 
viable if the holographic conjecture\cite{BoussoRMP} 
holds, i.e. if the
entire description of the world behind 
any
horizon
can be fully described on 
its
bounding surface.
It was argued in \cite{nature}
that the holographic conjecture is at best valid
when there is no trapped surface behind the horizon,
but it may otherwise in some sense be true. Whether
or not it is true, the fact remains that, for the observers who
remain confined to the ``outside" of the horizon,
the horizon entropy somehow captures the 
number of ways that the world beyond the horizon can 
affect the world outside. 

Black hole pair creation rates are sometimes
cited for a counter-argument to this viewpoint,
since the rates are proportional to $\exp{A/4}$
which is interpreted as a density of states factor
counting  degeneracy of {\it all} black hole states.
However it must be recalled that the instanton
method for computing a tunneling rate just
yields the probability for the most probable 
transition and the ones in its immediate vicinity.
Hence the degeneracy factor is not counting
{\it all} interior states of the black hole since 
these are beyond the scope of the instanton 
method. 

One clue to the nature of the horizon states
counted by $A/4$ comes from an
old analysis by Candelas and 
Sciama\cite{Candelas-SciamaPRL, Sciama}.
They showed how 
the relationship between near equilibrium transition rates 
for a system in contact with a horizon and horizon area
is extended to non-equilibrium processes.
They interpreted the viscous dissipation rate of a shearing horizon, via 
the fluctuation-dissipation theorem, in terms
of the quantum gravitational spectrum of shear fluctuations.
This explains ``why" a horizon has a coefficient of viscosity, and
suggests that it is, 
at least in part,
the quantum shear states of the fluctuating
horizon that the entropy counts.
This seems a good beginning, but it is surely not 
the whole story. In addition to the shear viscosity term
in Eq. (\ref{Ray}) for the horizon evolution there is
also a bulk viscosity ($\theta^2$) term. Morover, 
even when both these terms are
absent the horizon acts as a 
"perfect dissipator"\cite{Candelas-SciamaPRL,Sciama}
via just the area expansion. 

Another idea is that the horizon entropy arises from the entanglement of
short distance field fluctuations on either side of the horizon\cite{Sorkin}.
This notion is quite appealing since it traces the entropy directly to the
defining character of the horizon as a causal barrier that hides information,
and it naturally accounts for the scaling with area. Moreover, it allows
the generalized second law to be understood as a consequence of
causality\cite{SorkinGSL}. However, it is problematic in the
quantitative measure of missing information (which appears to be infinite in
ordinary field theory and to depend on the number of species) and in the
neglect of the quantum fluctuations of the horizon itself. These issues are
tied up with the renormalization of Newton's constant\cite{G},
though a completely satisfactory understanding at the statistical level is
far from being at hand. The finiteness problem is absent in the loop quantum
gravity approach\cite{Rovelli}, which can be viewed as a counting of intertwiners
that characterize the possible ways a spin network inside the black hole
could link to the one in the exterior. However, that approach remains to be
tied to the semiclassical limit in a quantitatively convincing manner.
Interesting perspectives on some of these issues are given in the review article by
Bekenstein\cite{BekMG7}.

While no definitive answer as to the 
ultimate nature of horizon entropy 
seems immediately at hand, 
an abundance of insight has been
gleaned from the three decades of
work. Perhaps the time is ripe
to synthesize this insight and make 
the leap to a new conception. 

\section*{\large ACKNOWLEDGMENTS}
We thank A. Ashtekar, P. Chru\'sciel, and G. Galloway for helpful correspondence.
This work was supported in part by the National Science Foundation
under grant No. PHY98-00967 and NATO Collaborative Linkage Grant CLG.976 417.


\edoc